\documentclass[aps,twocolumn]{revtex4}
\usepackage{amsfonts,amssymb,amsmath}            % for math symbols.
\usepackage[]{graphics,graphicx,epsfig}            % for graphics figures.
\usepackage{amsthm}                              
\def\identity{\leavevmode\hbox{\small1\kern-3.8pt\normalsize1}}

\newtheorem{lemma}{Lemma}
\newcommand{\ket}[1]{\left | #1 \right\rangle}
\newcommand{\bra}[1]{\left \langle #1 \right |}
\newcommand{\half}{\mbox{$\textstyle \frac{1}{2}$}}

\newcommand{\proj}[1]{\ket{#1}\bra{#1}}
\renewcommand{\epsilon}{\varepsilon}
\begin{document}

\title{The Computational Power of Symmetric Hamiltonians}
\date{\today}

\author{Alastair \surname{Kay}}
\affiliation{Centre for Quantum Computation,
             DAMTP,
             Centre for Mathematical Sciences,
             University of Cambridge,
             Wilberforce Road,
             Cambridge CB3 0WA, UK}
             
\begin{abstract}
The presence of symmetries, be they discrete or continuous, in a physical system typically leads to a reduction in the problem to be solved. Here we report that neither translational invariance nor rotational invariance reduce the computational complexity of simulating Hamiltonian dynamics; the problem is still BQP complete, and is believed to be hard on a classical computer. This is achieved by designing a system to implement a Universal Quantum Interface, a device which enables control of an entire computation through the control of a fixed number of spins, and using it as a building-block to entirely remove the need for control, except in the system initialisation. Finally, it is shown that cooling such Hamiltonians to their ground states in the presence of random magnetic fields solves a QMA-complete problem.
\end{abstract}

\maketitle

\section{Introduction}

Within the study of quantum computation, it is desirable to find `natural' problems for a quantum computer to solve i.e.~those that exhibit intrinsically quantum properties which elucidate the power of the device. For example, Feynman first proposed the quantum computer as a device that naturally simulates Hamiltonian dynamics. More recently, it has become apparent that finding the ground state energies of Hamiltonians is QMA-complete \cite{KSV02a, oliveira-2005,Kempe,gottesman,Kay:07} i.e.~this is a natural problem for the class where solutions can be efficiently verified on a quantum computer. The question that naturally arises with regards to both of these problems is that of the minimal properties that any such Hamiltonian must posses. The same question, viewed from another perspective demands when we should expect efficient classical approximations to the properties and dynamics of Hamiltonians. 

It is often found that the presence of symmetries, such as translational or rotational invariance, vastly simplify a problem. It may be hoped that accounting for these properties could make classical simulations of ground state, thermal or dynamic properties tractable and yet pertinent to physical phenomena. A straightforward example of this reduction is the description of a two-qubit mixed state $\rho$. To do this ordinarily requires 15 real parameters. However, if the state is rotationally invariant, i.e.~$(U\otimes U)\rho (U^\dagger\otimes U^\dagger)=\rho$ for all single-qubit unitaries $U$, then the family of possible states, the Werner states \cite{werner_state}, is parametrised by a single number. Similarly, under the discrete permutation symmetry, $\text{SWAP}\cdot\rho\cdot\text{SWAP}=\rho$, the family of states only contains 10 real parameters. The hope that the introduction of translational invariance reduces the complexity of finding ground states underpins studies such as Matrix Product States \cite{Whi92a,Whi93a,DMRG_period}.

In this paper, we are primarily concerned with the effects that these symmetries have on the ability to efficiently simulate Hamiltonian dynamics, and report that, in fact, the symmetries have no bearing; the problem is BQP-complete i.e.~as hard as any quantum computation is to simulate on a classical computer. We will show this by constructing Hamiltonians that implement arbitrary quantum computations.

The first steps towards incorporating translational invariance, for both Hamiltonian evolution and ground state properties, were taken in \cite{Kay:07}, which traded the spatial variation for another property such as the local Hilbert space dimension, which grew as $\text{poly}(N)$ for $N$ spins. Necessary to this construction was the inclusion of both a time label (clock) and position label at each site, so that all the information was available locally. Processing was then achieved by implementing a read head that moved backwards and forwards over the system data. Here, we use global control (GC) schemes \cite{Lloyd:1993a,Kay:thesis} to remove the necessity of both time and position labels. In Sec.~\ref{sec:2}, we will describe how to apply global commands through Hamiltonian evolution leading, in Sec.~\ref{sec:3}, to the realisation of a translational invariant, nearest-neighbor Hamiltonian of fixed local Hilbert space dimension that implements arbitrary quantum computations, thus implying the classical intractability of simulation of the dynamics, of which there is already some evidence \cite{norbert}. The construction is easy to motivate because a GC scheme typically works by repeated application of a finite set of pulses ``$A$", ``$B$" etc., which are local gates applied uniformly to all qubits in the system. As such, we immediately lose the need for spatial resolution in our Hamiltonian. To remove the time resolution, the program sequence is written on the initial state of some of the spins instead of encoding it in the Hamiltonian. Proceeding to a proof of QMA-completeness for ground state properties still requires the introduction of spatially varying magnetic fields (Sec.~\ref{sec:4}), but this in turn has severe implications for the cooling of physical systems in the presence of random external fields.

\section{A Universal Quantum Interface} \label{sec:2}

Lloyd and co-workers proposed the concept of a Universal Quantum Interface (UQI) \cite{UQI}. They proved that through manipulation of a single spin which is coupled to a larger system, the entire quantum system can be controlled, such that a quantum computation can be implemented on it. The proof was non-constructive, and there has recently been some interest in how such a device might be implemented \cite{Burgarth:07}. An example of one of the potential benefits of such a scheme is that it would in principle allow one to isolate the bulk system from the environment so that it's much less susceptible to noise. In comparison to schemes for Hamiltonian evolution, the UQI protocol also incorporates the ability to prepare the initial state of the system, although we will not explore that aspect here. Control theory proves the existence of solutions and offers numerical techniques to determine high-fidelity control \cite{Burgarth:07}. In contrast, we realise an exact UQI constructively. This will allow us to introduce most of the concepts required throughout the paper. Indeed, we will use the UQI as a building block in the design of our computing Hamiltonians.

\begin{figure}
\begin{center}
\includegraphics[width=0.45\textwidth]{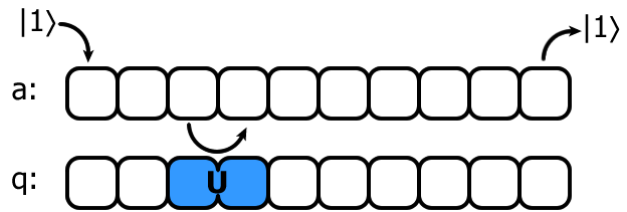}
\end{center}
\caption{The basic scenario of a universal quantum interface. At each site, there are 2 qubit systems, $a$ and $q$. All the $a$s are initialised as $\ket{0}$. A $\ket{1}^a$ is input at site 1, and its arrival at site $N$ is monitored. The hopping Hamiltonian is such that as the excitation moves between site $i$ and $i+1$, a unitary $U_{i,i+1}$ is enacted on the computational qubits $q$. Thus, upon arrival at site $N$, the unitary $U_{N-1,N}\ldots U_{1,2}$ has been implemented. Furthermore, this transfer is efficient.} \label{fig:uqi}
\end{figure}

Consider a linear chain of $N$ 4-dimensional spins. These 4-level spins can be assigned a structure as the tensor product of two qubits, labelled $a$ and $q$. Systems $q$ will contain computational qubits, and systems $a$ will be used as a `read head'. If the spins are coupled by an interaction
\begin{equation}
H_{\text{UQI}}=\half\sum_{i=1}^{N-1}(X_i^aX_{i+1}^a+Y_i^aY_{i+1}^a)\otimes S_{i,i+1}^q,	\label{eqn:uqi}
\end{equation}
where $S_{i,i+1}^q$ is the swap operation between computational qubits $i$ and $i+1$, and $X$ and $Y$ are the standard Pauli operators, then control of just spins 1, 2 and $N$ is sufficient to realise universal quantum computation. We can understand this by transforming the Hamiltonian into a state transfer system. We are interested in the subspace when all the $a$ qubits are in the state $\ket{0}$, except for qubit $i$ which is in the state $\ket{1}$. We choose to denote that as $\ket{i}^a\ket{\psi_i}^q$, where $\ket{\psi_i}^q$ is the state of the computational qubits in that step, and satisfies $\ket{\psi_i}=S_{i-1,i}\ket{\psi_{i-1}}$. With this definition, one can see that
$$
H_{\text{UQI}}\ket{i}^a\ket{\psi_i}^q=
\ket{i-1}^a\ket{\psi_{i-1}}^q+\ket{i+1}^a\ket{\psi_{i+1}}^q 
$$
($\ket{0}^a$ and $\ket{N+1}^a$ are taken to be $0$) which is just the same as the model of state transfer studied by Bose \cite{Bos03}, except that one needs to apply local magnetic fields on spins 1 and $N$. In particular, Bose studied the evolution $\ket{1}^a\ket{\psi_{1}}^q\rightarrow\ket{N}^a\ket{\psi_{N}}^q$ \cite{Bos03,Kay:2004c}, where one can achieve an arrival probability of $O(N^{-2/3})$ within a time $O(N)$ \cite{Bos03}. Moreover, this transfer is heralded -- by measuring system $a$ on qubit $N$, we know whether or not the transfer has occurred without disturbing the computational qubits. If the transfer is not finished, we simply wait and try again. With this strategy, one can achieve an evolution which fails with probability less than $\varepsilon$ in a time $O(M^{5/3}\log(1/\varepsilon))$ \cite{Bos03}. The setup is depicted in Fig.~\ref{fig:uqi}.

The computational scheme now proceeds as follows. We assume the state to be initialised with all the $a$ qubits in $\ket{0}$, and the computational qubits are in state $\ket{\psi_1}$. The read head is initialised by placing the $a$ qubit of spin 1 in the $\ket{1}$ state, and we perform a heralded transfer to spin $N$, which means that when it is observed that the excitation has arrived at $\ket{N}^a$, we reset the spin to $\ket{0}$, and the computational qubits have changed to $S_{N-1,N}\ldots S_{2,3}S_{1,2}\ket{\psi_1}^q$, which is simply a cyclic permutation of the computational qubits. This can be repeated, allowing any computational qubit to be placed on spin 1. Similarly, by starting the excitation on spin 2, and removing it from spin $N$, the computational qubits $2$ to $N$ undergo a cyclic permutation, allowing us to place any single qubit on spin 2 without disturbing spin 1. One we have any arbitrary pair of qubits on spins 1 and 2, which requires no more than $\half N$ cycles, our control of these two qubits allows the implementation of an arbitrary one- or two-qubit gate, which is sufficient for universal quantum computation.

\subsection{A Toolbox of UQI Gadgets}

Having demonstrated a relatively simple way to implement a UQI, we will now show how to modify the scheme to improve its capabilities. Our first observation is that we can replace the swap gate, $S^q_{i,i+1}$ in Eqn.~(\ref{eqn:uqi}) with any unitary $U_{i,i+1}^q$, and the transfer $\ket{1}^a_1\rightarrow\ket{1}^q_N$ implements the operation $U_{N-1,N}\ldots U_{2,3}U_{1,2}\ket{\psi_1}^q$. The only difference is that we must ensure that the Hamiltonian remains Hermitian, which we do by writing it as
$$
H_{\text{UQI}}^{(1)}=\sum_i(\sigma^{-a}_i\sigma^{+a}_{i+1})\otimes U_{i,i+1}^q+(\sigma^{+a}_i\sigma^{-a}_{i+1})\otimes U_{i,i+1}^{\dagger q}.
$$
The Hamiltonian remains translationally invariant provided $U_{i,i+1}$ is the same as $U_{1,2}$, just acting on different qubits.

The next step is to see how to implement $U_{i,i+1}$ only on every second qubit. We achieve this by increasing the dimension of system $a$ to 3. The idea is that instead of having a Hamiltonian where the read-head hops to the right, implementing $U$ as it goes, it will alternate its value between 1 and 2 as it hops, and will only implement $U$ if it's hopping from 1 to 2, and not 2 to 1. Thus, the new Hamiltonian reads
$$
H_{\text{UQI}}^{(2)}=\sum_i\ket{02}\bra{10}^a_{i,i+1}\otimes U_{i,i+1}^q+\ket{01}\bra{20}^a_{i,i+1}\otimes \identity_{i,i+1}^q+h.c.
$$
When we initialise the read-head on spin 1, whether it's in the state $\ket{1}$ or $\ket{2}$ determines which pairs of computational qubits the $U$ is applied to, pairs $(2i-1,2i)$ or $(2i,2i+1)$, and also what state we should be checking for on spin $N$. If $N$ is odd, the read-head undergoes an even number of steps and arrives in the same state it started on spin 1.

Finally, we might like the choice of applying more than one different $U$, say $U$ and $V$. Again, we can achieve this with the system $a$ having 3 levels.
$$
H_{\text{UQI}}^{(3)}=\sum_i\ket{01}\bra{10}^a_{i,i+1}\otimes U_{i,i+1}^q+\ket{02}\bra{20}^a_{i,i+1}\otimes V_{i,i+1}^q+h.c.
$$
By initialising the read-head on spin 1 as $\ket{1}^a$, the $\ket{1}^a$ hops from site to site, and as it hops it implements $U$. Similarly, initialising spin $1$ in the state $\ket{2}^a$ causes $V$ to be implemented.

These constructions are readily combined so that if we want to implement, say, two different global commands, each of which only acts on every second qubit, then we can construct a translationally invariant Hamiltonian that does it, using a read-head of dimension 5, and thus an overall spin dimension of 10. It is notationally convenient at this stage to decompose the 5-dimensional read-head at each site $i$ into
$$
(a_i\otimes r_i)\oplus n_i.
$$
The state $\ket{n}$ (the equivalent of $\ket{0}^a$ in the previous notation), is the state which all the read-head systems are initialised in, except for one, and is used to indicate the absence of the read head. The systems $a_i$ and $r_i$ are both qubits. $a_i$ indicates if the read-head is `active', and $r_i$ contains the `program information', i.e.~which of the two gates to implement. When we propagate the read-head to the right, the interaction that is implemented on the data qubits is conditioned on the value of the read-head, and whether the read-head is active. As the read-head hops, we flip the active setting, so that it only applies an operation on every second qubit.

\section{Computation by Hamiltonian Evolution} \label{sec:3}

In \cite{shepherd}, a global control scheme was developed based on just two nearest-neighbor gates, SWAP ($S_{i,i+1}$) and
$$
G_{i,i+1}=\identity\oplus(Z-iY)/\sqrt{2}.
$$
These gates need to be applied to distinct qubit pairings $(2i-1,2i)$ and $(2i,2i+1)$ across the entire lattice. Evidently, this ties in very usefully with our UQI construction. If we implement these two gates within the Hamiltonian, then the 10 level system is capable of universal quantum computation, where we only need single-spin control of spins 1 and $N$ to perform the transfer of the read-head, and to specify the program sequence (i.e.~the order in which global commands are implemented).

\begin{figure}
\begin{center}
\includegraphics[width=0.45\textwidth]{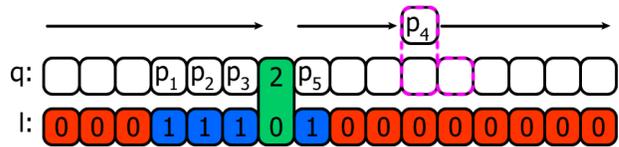}
%\vspace{-0.5cm}
\end{center}
\caption{Schematic of the Hamiltonian's mechanism. At each site, there is a label to specify if that site is a computational qubit or program qutrit. One location in the program bus is marked as the active region, and that value is stored in the read-head qubit $r_i$. The read-head applies a unitary to each computational qubit consecutively, controlled off its value. When it reaches the active program region, it exchanges its data with the next step in the program.} \label{fig:schematic2}
%\vspace{-0.5cm}
\end{figure}

Our aim is now to construct a Hamiltonian that implements the entire evolution of the computation without even this basic level of control, but instead retaining the ability to prepare the system in some initial state. The first step is to remove the asymmetry between operations at either end of the chain; we introduce periodic boundary conditions and a special marker state. The read-head will be emitted from one side of the marker state, and will arrive at the other side. It will then be down to our preparation of the initial state to select where this marker state is. The second step is to incorporate the program sequence. Again, we will do this by writing the additional information in the initial state. This will require an additional two-level system at each site to denote whether that site constitutes a computational qubit or a spin that holds the program data. The previously mentioned marker will then be used to denote which of the program spins is the one that's being actively implemented. Again, the read-head will be emitted from one side of this marker, but when it arrives on the other side, it will move the marker onto the next program spin (see Fig.~\ref{fig:schematic2}). The natural start and end points to the computation are when the active program marker is at either end of the program bus, and are detected by a change in the $l_i$ label between program and data spins.

To be specific, consider a 1D chain of spins of local dimension 31, which can be decomposed into several subsystems
\begin{equation}
s_i=(q_i\otimes l_i\otimes ((a_i\otimes r_i)\oplus n_i))\oplus m_i.	\label{eqn:hilbert_space2}
\end{equation}
The system $q_i$ is a qutrit, serving two different purposes depending on the label of the qubit system $l_i$. If the label is $\ket{0}$, then $\ket{0}^q$, $\ket{1}^q$ encode a computational qubit (the system $q$ in the UQI construction), otherwise the qutrit $q_i$ contains program information -- ``skip'', $G$ or $S$. There is a single state which is not used yet, $\ket{2}^q\ket{0}^l$, and is reserved as the marker to denote which program trit is active. The single level $m_i$ is used to help moving the marker over a `skip' label.

Whether a particular global gate works on pairings of qubits $(2i-1,2i)$ or $(2i,2i+1)$ is solely determined by the alignment of the active program trit with respect to the start of the block of data qubits, which is why we require skip; such that the relative alignment changes. Assuming that there are an odd number of spins in the system, if a read-head leaves a program trit in the active state, it returns in the inactive state. Consequently, an inactive read-head in the location of the active program trit can be used to indicate that the read-head should move the active trit to the next one.

The main term in the Hamiltonian $H_{prop}$ involves read-head propagation, and comes directly from the UQI construction. As such, we will not repeat it here. We must additionally incorporate a term to stop the read-head propagating if the state $\ket{2}^q\ket{0}^l$ is present i.e.
\begin{eqnarray}
&\tilde H_{prop}^i=H_{prop}^i(\identity-\proj{20}^{ql}_i)+&	\nonumber\\
&\sum_x\ket{n20}\bra{x120}^{raql}_i\otimes(\ket{x0}^{ra}\bra{n}\otimes\identity^{ql})_{i+1}+h.c.& \nonumber
\end{eqnarray}
Next, program manipulation (when the read-head is in the neighborhood of the currently active program state):
\begin{eqnarray}
H_{prog}^i&=&\sum_{x,y}(\ket{n}\bra{x0}^{ra}\otimes\ket{x+1}\bra{2}^q\otimes\ket{1}\bra{0}^l)_i	\nonumber\\
&&\!\!\!\!\otimes(\ket{y1}^{ra}\bra{n}\otimes\ket{2}\bra{y+1}^q\otimes\ket{0}\bra{1}^l)_{i+1}+h.c.	\nonumber
\end{eqnarray}
except that this doesn't (yet) handle the skip. First, if we're in a region where we've just arrived back from doing a loop, and should be moving onto a skip label
\begin{eqnarray}
H_{s1}^i&=&\sum_x(\ket{n}\bra{x0}^{ra}\otimes\ket{x+1}\bra{2}^q\otimes\ket{1}\bra{0}^l)_i	\nonumber\\
&&\otimes(\ket{m}\bra{n}\bra{01}^{ql})_{i+1}+h.c.	\nonumber
\end{eqnarray}
and, secondly, the step over the skip label
\begin{eqnarray}
H_{s2}^i&=&\sum_x(\ket{n}\ket{01}^{ql}\bra{m})_i\otimes	\nonumber\\
&&(\ket{x1}^{ra}\bra{n}\otimes\ket{2}\bra{x+1}^q\otimes\ket{0}\bra{1}^l)_{i+1}+h.c.	\nonumber
\end{eqnarray}
Finally, the total Hamiltonian is
\begin{equation}
H_T=\sum_i\tilde H_{prop}^i+H_{prog}^i+H_{s1}^i+H_{s2}^i,	\label{eqn:ham_final}
\end{equation}
which is entirely a sum of two-body terms, so it can be represented as $\sum_{i}h_{i,i+1}$. Moreover, for the permutation operator
$$
P=\sum_{i_1\ldots i_N=0}^{30}\ket{i_1i_2\ldots i_N}\bra{i_2i_3\ldots i_Ni_1},
$$
we have that
$$
PH_TP^\dagger=H_T
$$
i.e.~the Hamiltonian is translationally invariant.

If there are $N$ computational qubits in the system, then any efficient quantum algorithm is described by $\text{poly}(N)$ bits, and the total number of qubits in the system is $M$, the sum of these two. Again, we can invoke the fact \cite{Bos03} that after an evolution time $O(M)$, the probability of having successfully completed the computation of $O(M^{-2/3})$, which is an efficient implementation. So, this translationally invariant, nearest neighbor Hamiltonian evolution can implement any quantum computation, starting from a separable state. The fact that it starts from a separable (although not translationally invariant) state is important since it ensures that we are not encapsulating the difficulty of the problem within the preparation of the initial state. We conclude that the problem is BQP-hard. However, it is also known how to simulate Hamiltonian evolution on a quantum computer \cite{mick}, so the problem is BQP-complete. Thus, as strongly as we believe that quantum computation is more powerful than classical computation is how strongly we believe that simulation of Hamiltonian dynamics, even under the translational invariant restriction, is hard to simulate on a classical computer.

\subsection{Rotational Invariance}

\begin{figure}
\begin{center}
\includegraphics[width=0.45\textwidth]{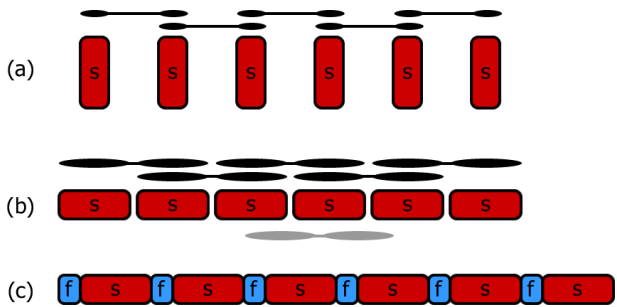}
\end{center}
\caption{(a) Schematic depiction of a 1D array of 31 dimensional spins (s). The black regions denote the nearest-neighbor interactions of the Hamiltonian. (b) The same system made rotationally invariant by encoding the states of the spin in logical states on a block of qubits. The logical spins, and the periodic Hamiltonian interactions are depicted. To make the system translationally invariant requires the inclusion of terms (gray) where the interactions do not align with the logical qubits of the state. (c) To regain translational invariance in the Hamiltonian, we introduce a flag state (f) before each block of logical qubits, denoting the start of that block.} \label{fig:ham_conversion}
\end{figure}

So far, we have seen how a 1D Hamiltonian with fixed local Hilbert space dimension and fixed range interactions, which is translationally invariant, can implement an arbitrary quantum computation. However, this discrete symmetry is not nearly as restrictive as the continuous symmetry of rotational invariance, which requires the Hamiltonian to satisfy
$$
U^{\otimes M}HU^{\dagger \otimes M}=H
$$
for all single qubit unitaries $U$. 
We will now show how to build this into the Hamiltonian, retaining translational invariance and the ability to perform arbitrary quantum computations. The first step is to take the 31 dimensional construction $H_T$, Eqn.~(\ref{eqn:ham_final}), and replace each spin with 10 qubits. Between these 10 qubits, there are several decoherence-free subsystems \cite{rob:revmod}. For an $N$ qubit system ($N$ even), there are decoherence-free subsystems which have $\binom{N}{N/2}(2j+1)/(N/2+j+1)$ levels for $j=0\ldots N/2$. These subsystems enable the storage of quantum information in a way that is not affected by collective decoherence $U^{\otimes N}$. Thus, encoding within one of these subsystems stores the information in a rotationally invariant way. We select any one of the 4 subsystems ($N=10$) that is large enough to encode the 31 levels. Transcribing $H_T$ into this new form automatically makes it rotationally invariant, although instead of being translationally invariant, it is periodic, with a repetition length of 10 qubits. We can thus write it as
$$
H_R=\sum_{i=1}^{M-1}h_{(10(i-1)\ldots 10i-1),(10i\ldots 10(i+1)-1)},
$$
denoting the blocks of logical spins. A summary of this is depicted schematically in Fig.~\ref{fig:ham_conversion}.
Note that the initial state that the Hamiltonian evolution acts on must also be encoded. However, it is encoded into fixed sized blocks which are separable from each other. Thus, the initial state can still be efficiently represented on a classical computer, so we have not transformed the problem of simulation into the preparation of the initial state -- it's still contained within the Hamiltonian evolution.

In order to reintroduce translational invariance, we want to incorporate a local patterning of states that enables us to detect the alignment of the blocks of qubits (Fig.~\ref{fig:ham_conversion}(c)). The technique that we use is much clearer if we concentrate, initially, on restricting the arbitrary rotations to rotations about a single axis, $U=e^{i\theta Z}$. Our problem is that a translationally invariant Hamiltonian will be made up of sums of terms, each of which acts on a block of qubits, comprised of two logical spins. We need to make sure that if a Hamiltonian term is not perfectly aligned with the block-wise patterning of the initial state, then it does not contribute to the evolution. To do this, we introduce a patterning of the qubits which is still rotationally invariant, and yet flags the start of each block of spins. For $Z$-rotation invariant states, this can be done by taking each set of qubits that constitutes a logical spin, and introducing a qubit in the $\ket{0}$ state between each of them. This means that in the initial state, there is never a pair of neighboring qubits in the $\ket{11}$ state. Thus, we use this to flag that a block of logical qubits is starting. So, each logical spin now constitutes 23 qubits, of the form $\ket{110q_10q_20q_30q_40\ldots q_{10}0}$ where $\ket{q_1q_2\ldots q_{10}}$ was the previous logical spin encoded into a decoherence-free subsystem, and the Hamiltonian is of the form $\proj{11}_{1,2}h_{(4,6,8\ldots 22),(27,29\ldots 45)}$.

\begin{figure}
\begin{center}
\includegraphics[width=0.45\textwidth]{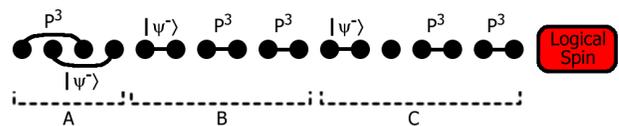}
\end{center}
\caption{The rotationally invariant state that flags, and the Hamiltonian that detects the flag, at the start of the block of 10 qubits encoding a logical spin within a decoherence free subsystem. Section A makes sure that when the Hamiltonian and herald state are offset by a single qubit, the overlap is 0. Section B, where the states $P_3$ are repeated 8 times, handles a relative shift of an even number of qubits. Section C does the same for an odd number of qubits, and the states $P_3$ are repeated 9 times. The difference in the number of repetitions of $P_3$ handles an edge effect that arises otherwise.} \label{fig:heralding}
\end{figure}

For $Z$-rotations, we have the advantage that the states $\ket{0}$ and $\ket{1}$ are rotationally invariant (but not superpositions of them). For arbitrary rotations $U$, the construction of suitable flag states is much more involved, and is based on the fact that $\ket{\psi^-}=(\ket{01}-\ket{10})/\sqrt{2}$ is a rotationally invariant two-qubit state. Hamiltonian terms can be constructed which detect the presence, $P^1=\proj{\psi^-}$, or absence, $P^3=\identity-\proj{\psi^-}$, of such a state. There will be a correspondence between the Hamiltonian term that detects the flag state, and the flag state itself, the only difference is that if the Hamiltonian includes a term $P^3$, the flag state can be made out of any two-qubit pure state orthogonal to $\ket{\psi^-}$, such as $\ket{00}$. We proceed by observing that it is relatively easy to suppress misalignments between the flag state and the Hamiltonian term when the misalignment is by an even number of qubits, one simply ensures that a $\ket{\psi^-}$ in the flag state and a $P^3$ in the Hamiltonian align. This trick can be repeated for a misalignment by an odd number of qubits greater than 1. There are two concerns remaining. Firstly, whether there any edge effects arising and, secondly, how to deal with an offset of just one qubit. The first concern is overcome simply by using a large enough flag state, which is larger than the 10 qubits in the logical spin. An offset of one qubit is handled by incorporating four additional qubits in the flag state in the form $P^1_{1,3}\otimes P^3_{2,4}$. The interleaving of the two projectors ensures overlap after just a single shift. The entire state is depicted in Fig.~\ref{fig:heralding} and requires 43 qubits in total. The projector onto the state is written as
\begin{eqnarray}
P^{\text{flag}}_{1\ldots 43}&=&P^1_{1,3}\otimes P^3_{2,4}\otimes P^1_{5,6}\otimes\bigotimes_{i=0}^7P^3_{2i+1,2i+8}\otimes	\nonumber\\
&&\otimes P^1_{23}\otimes\identity_{24}\otimes \bigotimes_{i=0}^8P^3_{2i+25,2i+26}.	\nonumber
\end{eqnarray}
Thus, the overall translationally and rotationally invariant Hamiltonian acts on blocks of 106 qubits which are local on a 1D lattice. It is of the form
\begin{eqnarray}
H_{RT}&=&\sum_{i=1}^{M-1}P^{\text{flag}}_{53(i-1)\ldots 53(i-1)+42}\otimes P^{\text{flag}}_{53i\ldots 53i+42}\otimes	\nonumber\\
&&\otimes h_{(53(i-1)+43\ldots 53i-1),(53i+43\ldots 53(i+1)-1)},	\nonumber
\end{eqnarray}
where the $h_{(),()}$ are the same as in $H_R$. The majority of the cost in terms of the range of the Hamiltonian terms is due to the flag state, which we have made little attempt to optimise; the important element is that the range of the terms is independent of $N$.

\section{Ground State Properties in the Presence of Magnetic Fields} \label{sec:4}

In previous studies, Hamiltonian evolution has been used as a basis for classifying the problem of finding ground state energies of Hamiltonians with similar properties as QMA-complete. In the present case, this is not expected to be possible as there seems to be no way to encode the verifier's computation while retaining translational invariance. However, by breaking the translational invariance of $H_T$, one arrives at similar results to \cite{gottesman}, but only using local magnetic fields. To achieve this, we need to add several energy penalties; to detect the solution to the verifier circuit, to initialise ancillas in $\ket{0}$ and to prepare the initial state of the program tape for a specific computation corresponding to the verifier of the QMA problem. To implement these penalties, we need to be able to locally detect that we are either at the beginning or end of a computation, requiring an increase in the local Hilbert space dimension, such that the system $n$ has 2 levels. $\ket{0}^n$ will be used as before, to indicate that the read-head is inactive. $\ket{1}^n$ is a program command that will only get used once, as the first command. It is not necessary to program it, because two-body terms can readily detect the transition between the data and program spins. A further change is that the system $a$ must be increased to dimension 3, leaving the overall Hilbert space dimension as 49. The extra level in $a$ serves a dual purpose. Firstly, it can be used in the same way as $\ket{1}^n$, but to indicate the end of the computation, such that we can penalise the output qubit. Secondly, it is used to help ensure that the correct computational sequence occurs. We will adapt the Hamiltonian propagation such that if a read-head in either $\ket{0}^a$ or $\ket{1}^a$ arrives at the region of transition from data to program spins, it is converted into $\ket{2}^a$, which will continue to propagate through the program region until it gets to the active program spin, where it releases its information and gets reinitialised in $\ket{1}^a$. If the read-head reaches the end of the program region in the $\ket{2}^a$ state, it is deactivated, and the computation ends. In particular, this means that if the system were to be initialised without an active program label, the computation is much shorter than it would otherwise have been. The Hamiltonian is readily revised to take these alterations into account. An energy penalty for when the read-head passes a particular qubit in either $\ket{1}^n$ or $\ket{2}^a$ behaves exactly like the initial and final penalties that we require, so we simply use penalties
$$
H_{in}=\sum_i\proj{1}^n_i\otimes\proj{\bar x_i}_i^q
$$
where the tape value should be $x_i$ ($x_i=0$ for ancillas), and hence ${\bar x_i}$ implies a sum over all other possible program (data) states, including the active label $\ket{0}^l\ket{2}^q$. The final result term (to test the output of the verifier on qubit $o$) is similar:
$$
H_{out}=\proj{2}^a_o\otimes\proj{0}^q_o.
$$
We have to be sure that the computation is initialised correctly, with all the spins correctly arranged. This is achieved by adding a constant term $H_b$. On program spins, this term is $\proj{0}^l\otimes(\proj{0}+\proj{1})^q$, ensuring  that they are never data qubits. On data spins, this term is the opposite, $\proj{1}^l\otimes\identity^q+\proj{0}^l\otimes\proj{2}^q$. 
Taking all of this into account, one can directly apply the projection lemma of \cite{Kempe},
\begin{lemma}
Let $H=H_1+H_2$ be the sum of two Hamiltonians operating on some Hilbert space $\cal H=\cal S+\cal S^\perp$. The
Hamiltonian $H_2$ is such that $\cal S$ is the ground state eigenspace (with eigenvalue 0) and the eigenvectors in $\cal S^\perp$ have eigenvalue at least $J> 2\|H_1\|$. Then,
 $$\lambda(H_1|_{\cal S}) - \frac{\|H_1\|^2}{J-2\|H_1\|} \le \lambda(H).$$
$\lambda(H)$ denotes the smallest eigenvalue of $H$.
\end{lemma}
For example, with $J=8\|H_1\|^2+2\|H_1\|$, one obtains
%In particular, one can select, say $J=8\|H_1\|^2+2\|H_1\|$ in order to provide the bounds
$\lambda(H_1|_{\cal S}) - \frac{1}{8} \le \lambda(H).$
We start with the total Hamiltonian
$$
H_{total}=-J_0H_T+J_bH_b+J_{in}H_{in}+J_{out}H_{out}+\kappa\identity,
$$
where $\kappa=-J_0\lambda(-H_T)$. $H_T$ takes a slightly different form to standard proofs since instead of mapping to a Heisenberg chain, it maps to an XX model, and hence the eigenvalues are $2\cos(\pi m/(M+2))$ for $m=1\ldots M+1$. There is an energy gap for computations that are fewer than $M$ steps, as well as a gap to the first excited state,
\begin{eqnarray}
-2\cos\left(\frac{\pi}{M+1}\right)+2\cos\left(\frac{\pi}{M+2}\right)&\geq&\frac{c}{M^2}=\Delta E	\nonumber\\
-2\cos\left(\frac{2\pi}{M+2}\right)+2\cos\left(\frac{\pi}{M+2}\right)&\geq&\Delta E	\nonumber
\end{eqnarray}
for some constant $c>0$. 
In the case where `yes' solutions exist, $\lambda(H_{total})=J_0\lambda(-H_T)+\kappa=0$. In the case where there are only `no' solutions, we assign $H_2=-J_0H_T+J_bH_b+\kappa\identity$ to find that
$
J\geq\min\left(J_b,J_0\Delta E\right),
$
and, furthermore,
$
\lambda(H_{total})\geq \lambda(H_1|_{{\cal S}_0})-1/8,
$
provided $J\geq 8(J_{out}+J_{in})^2+2(J_{out}+J_{in})$, imposing a polynomial relation between $J_0,J_b$ and $J_{in},J_{out}$. Repeating the process on $\lambda(H_1|_{{\cal S}_0})$ with $H_1'=J_{out}H_{out}|_{{\cal S}_0}$, shows that provided $J_{in}\geq 8J_{out}^2+2J_{out}$,
$$
\lambda(H_1|_{{\cal S}_0})\geq \frac{J_{out}(1-\varepsilon)}{M+1}\sin^2\left(\frac{(M+1)\pi}{M+2}\right)-\frac{1}{8}
$$
where the verifier circuit of our QMA problem accepts the result with probability less than $\varepsilon$. Thus, by selecting 
$
J_{out}=(M+1)\sin^{-2}\left(\frac{\pi}{M+2}\right)\leq c'M^3,
$
all the terms $J_{out},J_{in},J_b$ and $J_0$ are polynomial in $M$, and
$$
\lambda(H_{total})\geq \frac{3}{4}-\varepsilon.
$$
Distinguishing the ground state energy of this Hamiltonian to within $1/\text{poly}(M)$ determines the existence of `yes' solutions, and thus finding the ground state energy is QMA-complete. In comparison to \cite{gottesman}, all of the spatially varying terms are local magnetic fields. An identical proof holds using $H_{RT}$, our qubit Hamiltonian which is both translationally and rotationally invariant, although the penalties are no longer local magnetic fields and are, instead, 53-body.

\section{Conclusions}
Making use the GC scheme introduced in \cite{shepherd}, we have developed three main results. Firstly, the Universal Quantum Interface was described, and used as a building block for the second part, which showed that the evolution of a fixed Hamiltonian which is translationally invariant on a nearest-neighbor chain and has fixed spin dimension can simulate any arbitrary quantum computation, thereby suggesting that the evolution is hard to simulate classically because it is a BQP-complete problem. Even simulations over short time scales, $O(\Delta E^{-1/4})$, reveal the solution since we can project onto the heralded outcome. We have also extended this result to include qubit Hamiltonians which are translationally and rotationally invariant, and still act on $O(1)$-nearest neighbors which are local on a 1D lattice.

Finally, finding the ground state of a translationally invariant Hamiltonian in the presence of a specific sequence of local magnetic fields is QMA-complete. This has a more useful interpretation as a specific example of a random magnetic field i.e.~finding the ground state of a translationally invariant Hamiltonian on $M$ spins in the presence of a random local magnetic field of size $O(1/M^2)$ is QMA-complete. This has some important consequences for physical scenarios involving cooling. For example, if one were to couple a refrigerator to a quantum system in an attempt to cool it, if the coupling is too strong, it would take a prohibitively long time to reach the ground state of the system, and that ground state is not the same ground state when it's not coupled to the refrigerator \footnote{Although in the particular example that we have constructed, the ground state of the coupled system has significant overlap with the degenerate ground state space of the original Hamiltonian.}. Conversely, the weaker one couples the system to the refrigerator, the longer it takes to cool. This helps to provide a motivation for the use of topological and self-correcting systems, which are carefully designed such that local magnetic fields cannot affect the system degeneracy \cite{DKLP02a}.

Since completing this work, we have been made aware of related work on evolution by translationally invariant Hamiltonians \cite{karl} and subsequent work \cite{followup} which reduces the local Hilbert space dimension in that case. This could presumably be used to reduce the number of levels required for the other results presented in this paper.

In the future, we intend to examine whether the present work enables any useful insights into the problem of determining ground states for translationally invariant systems. It certainly leads to some natural conjectures which we are working to prove. Also, the technique for converting translationally invariant Hamiltonians into translationally and rotationally invariant Hamiltonians may be usefully applied to show that the two ground state problems have the same computational complexity -- it is certainly true that the eigenstates of $H_T$ map through to $H_{RT}$. It remains to prove that the ground state of $H_T$ maps to the ground state of $H_{RT}$.

AK would like to thank F. Verstraete for useful conversions and Clare College, Cambridge for financial support.

\end{document}